\documentclass[twocolumn,prl]{revtex4}
\pdfoutput=1 
\usepackage{graphicx,wrapfig}

\usepackage{color,mathptm,amsmath,bm}
\begin{document}

\thispagestyle{empty}
\bibliographystyle{unsrt}

\title{Non-specific DNA-protein interaction: Why proteins
can diffuse along DNA
%Why Shape Complementarity Induces an Osmotic Repulsion Between Protein and DNA
%Why Shape Complementarity Induces an Electrostatic Lock and Key Mechanism  Between Protein and DNA
}

% COMPILER AVEC PDFLATEX\\\\

\typeout{Output format A4.}
\textwidth=19cm
\oddsidemargin=-1.5cm
\evensidemargin=-6cm
\topmargin=-2.3cm
\headheight=1.5cm
\headsep=1cm
\textheight=24.cm
\parindent=0.5cm
\newfont{\itg}{cmbxti10 at 10pt}
\newcommand{\anf}{\sffamily}

\author{Vincent Dahirel$^1$, Fabien Paillusson$^{2,3}$, Marie Jardat$^1$, Maria Barbi$^{2,3}$, Jean-Marc Victor$^{2,3}$}

\affiliation{$^1$UPMC Univ Paris 06, UMR 7195, PECSA, F-75005 Paris, France,
 $^2$UPMC Univ Paris 06, UMR 7600, LPTMC, F-75005 Paris, France, 
 $^3$CNRS, UMR 7600, F-75005 Paris, France}

\begin{abstract}
%The shape complementarity of DNA Binding Proteins and
%their specific DNA sequence enables to maximize the number of direct
%interactions with DNA basepairs.
The structure of DNA Binding Proteins enables 
a strong interaction with their specific target site on DNA.
However, recent single molecule experiment reported that proteins can diffuse on
DNA. 
This suggests that the interactions between proteins and DNA play a role during the target 
search even far from the specific site.
It is unclear how these non-specific interactions optimize
the search process, and how the protein structure comes into play.
Each nucleotide being negatively
charged, one may think that the positive surface of DNA-BPs
should electrostatically collapse onto DNA. Here we show by means of
Monte Carlo simulations and analytical calculations that a counter-intuitive repulsion between
the two oppositely charged macromolecules exists at a nanometer range.
We also show that this repulsion is due to a local increase of
the osmotic pressure exerted by the ions which are trapped at the interface.
For the concave shape
of DNA-BPs, and for
realistic protein charge densities, 
we find that the repulsion
pushes the protein in a free energy minimum at a distance from DNA. 
As a consequence, a favorable path exists along which proteins can slide 
 without interacting with the DNA
bases.  
When a protein encounters its target, the osmotic barrier is
completely counter-balanced by the H-bond interaction, thus enabling
the sequence recognition.
\end{abstract}

\maketitle

DNA stores the genetic material of all living cells and viruses. This
huge amount of information is effective only if DNA binding
proteins (DNA-BPs) manipulates DNA in very specific locations.
When the protein finds its DNA target, the shape complementarity of DNA Binding Proteins and
their specific DNA sequence enables to maximize the number of hydrogen
bonds, thus leading to a strong protein-DNA association \cite{Jon99,Jan99,von07,Tak92,Via00,Kal04}.
The rate of protein-DNA association is however not controlled 
by the association step itself, but by the whole searching process.
It is well established now that DNA-BPs diffuse along DNA
before they reach their specific site \cite{revue2008}. 
During this search, the only interactions between protein and DNA 
which can play a role are non
sequence-specific.
Those non-specific interactions between protein and DNA remain poorly
documented. Altough the predominance of electrostatics is unquestionable
\cite{Jon99,Jan99,Tak92,von07,Via00,Kal04}, 
it remains unclear how the protein structure comes into play
\cite{Via00,Kal04,revue2008}. Does the typical concavity of DNA-BPs
which favors the specific association also influence the non-specific electrostatic interaction?
In DNA-protein complexes, the mean charge of the protein residues
located at the interface is positive \cite{Jon99,Jan99}.
Nevertheless, structural studies of non-specific complexes have shown
that the protein atoms and the DNA atoms are weakly packed together at
the interface \cite{Jon99,Jan99,von07,Via00,Kal04}, thus suggesting
that a force counterbalances the electrostatic attraction. 
In this letter, our purpose is to
establish the general mechanisms that control the mean force between
protein and DNA and that are applicable to a wide variety of
DNA-BPs. That goal in mind, we design coarse-grained DNA and protein models, rather
than detailed atomic models and investigate their interactions. 
First, we prove that
a short range repulsion exists 
when the shape of the protein is complementary to the shape of DNA.  
Second, we show that this repulsion increases when the protein
charge decreases, and we unravel the underlying physical mechanism. Finally, we
discuss in detail why this phenomenon is relevant to real biological
systems, thanks to statistical data of the protein charge and of the number of
H-bonds between protein and DNA.
 
\begin{figure}[!t]
\begin{center}
\includegraphics*[scale=0.44]{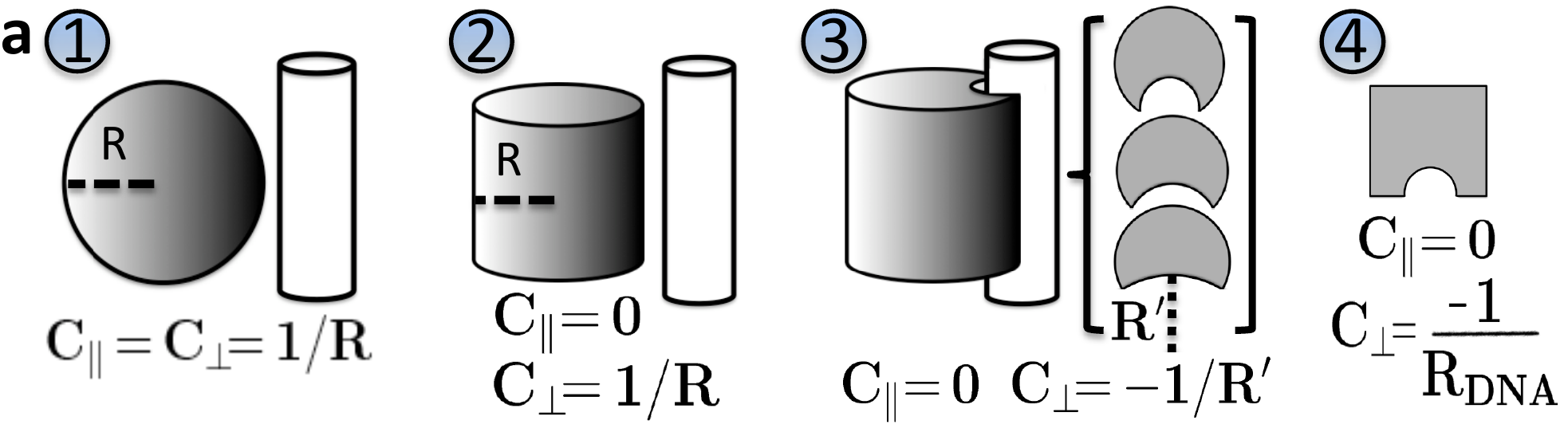}
\includegraphics*[scale=0.41]{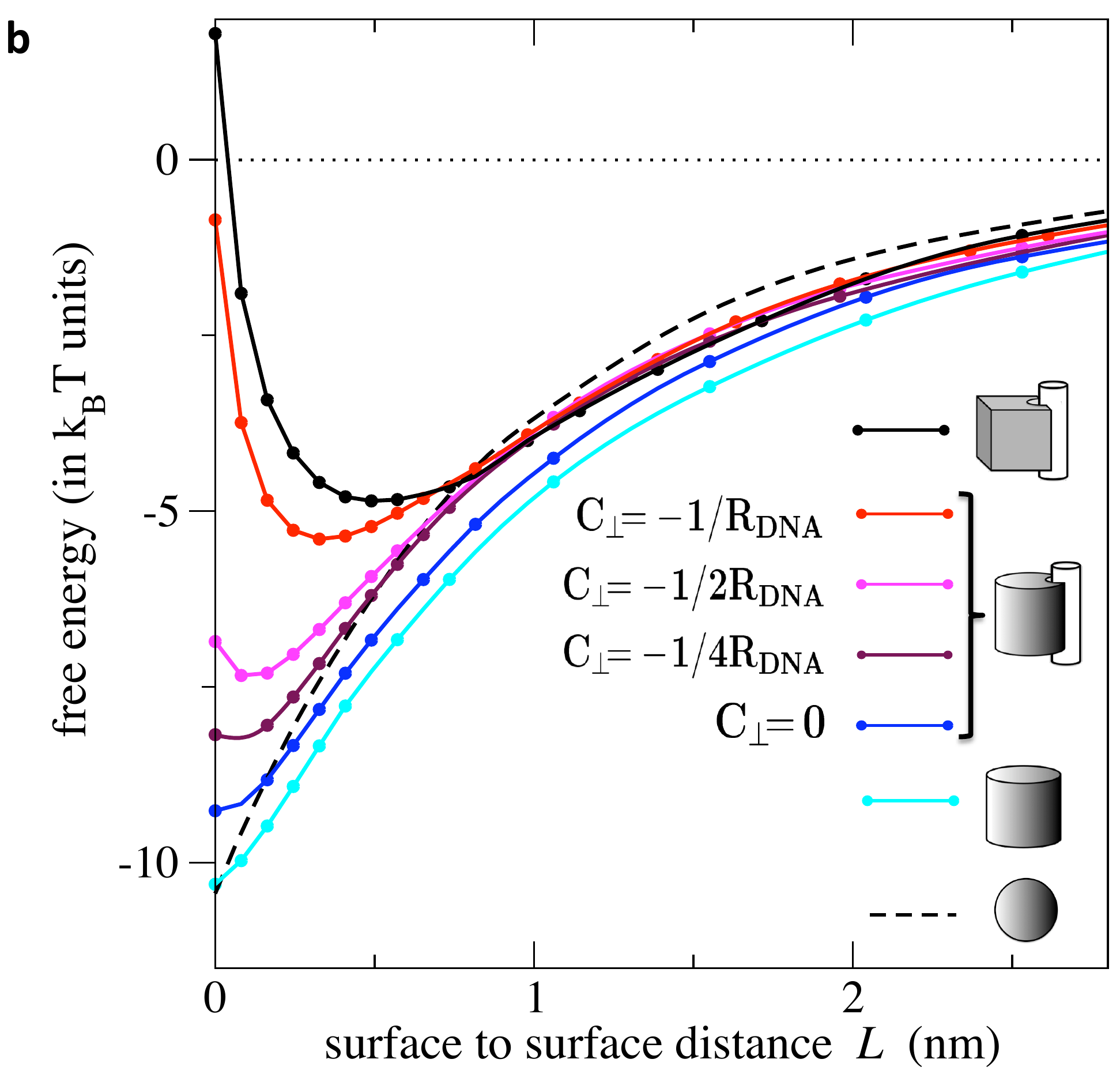}
\caption{{\bf Influence of the protein shape on the interaction. a,} Schematic view of
the model proteins. 
The height and diameter of the cylindrical proteins (2,3) are both
$5$~nm, as well as the side of the cubic protein (4) and the radius of
the sphere. The hollow cylindrical proteins (3) have a cylindrical
cavity, of curvature $C_\perp = 0, -0.25, -0.5$ or $-1$
nm$^{-1}$. {\bf b,} Free energy of the DNA-protein systems computed by
MC simulations.  The protein and DNA are immersed in a monovalent salt
whose Debye length $\lambda_D=$1~nm \cite{Note2} corresponds to
physiological conditions. The standard deviation of the free
energy is~$0.2$~k$_{\rm B}$T.}
\label{fig1}
\end{center}
\end{figure}

The most characteristic aspect of DNA-BPs is their shape
complementarity with DNA. As a matter of fact, the concave DNA-BPs can
cover the convex DNA with up to 35\% of their surface \cite{Jon99}. At
close contact, those interface regions exclude the solvent molecules
and form numerous weak bonds with DNA (mainly H-bonds
\cite{Jon99}). In a first instance, we artificially switch off these
H-bond interactions.  To probe the influence of protein shape in
controlling the non-specific electrostatic interaction, we monitor
changes in the potential of mean force upon modifying the curvature of
smooth model proteins along the DNA direction (noted $C_\parallel$)
and in the perpendicular direction ($C_\perp$) (see Fig.~\ref{fig1}a).
The charge of all model proteins is given by a single $+5e$ site
placed $0.7$~nm under the protein surface facing DNA.  The direct
electrostatic force in vacuum is therefore the same for any protein
shape investigated here. The DNA is modelled as a hard cylinder with
divalent charged sites. The water and the electrolyte ions are
described by the primitive model of electrolyte solutions
\cite{Han00}. This model has already been used to explain the less intuitive trends of
electrostatic interactions in solution, {\it e.g.} the attraction
between like-charged particles \cite{All98}, or the repulsion
between charged and neutral ones \cite{Dah08}. 
The relative permittivity of water $\epsilon_r$ is taken equal to $78.25$, and
the radius of the salt ions is $0.15$~nm.

The potential of mean force between a protein and a DNA molecule
separated by a distance $L$ is equal to the free energy of the global
system (protein, DNA and ions in water). At a fixed surface-to-surface
distance $L$, this energy only depends on the ion distribution. We
compute thus the free energy thanks to canonical Monte Carlo (MC) simulations
that sample the ion configurations \cite{DahirelPRE07,DahirelJCP07}.
We voluntarily freeze
the rotational degrees of freedom of the protein, and study the
interaction for the most attractive orientation, when the protein
cavity points toward DNA.  Indeed, this orientation is the one always
observed for specific and non-specific complexes, and we observed that
the free energy gets abruptly more repulsive when the protein rotates.
The protein and DNA are placed in a parallelepipedical
simulation box (275x275x150 nm) with periodic boundaries.
The results are reported in Fig.~\ref{fig1}b. 

The curvature
$C_\parallel$ slightly influences the range of the interaction, as
illustrated by the comparison of spherical and cylindrical proteins.
The effect of the curvature $C_\perp$ is remarkably more pronounced.
The free energy as a function of $L$, which is monotonic for $C_\perp
> 0$, becomes non-monotonic for $C_\perp < 0$ and exhibits then a
minimum $F_{\rm min}$ at a distance $L_{\rm min}$. For $L<L_{\rm
min}$, there is an unexpected repulsive free energy barrier between
the oppositely charged bodies, that reaches $\sim5$~k$_{\rm B}$T in
the case of perfectly matching surfaces ($C_\perp=-1/R_{\rm DNA}$).
%The concavity of DNA-BPs produces
%thus two opposing effects: It maximizes interatomic attractions, such
%as H-bonds, whereas it provokes a global repulsion of electrostatic
%origin. 
This behavior is weakly influenced by the shape of the remaining
surface of the protein: $F_{\rm min}$ varies from {\em e.g.} 
$-4.9$~k$_{\rm B}$T with a cubic protein to $-5.4$~k$_{\rm B}$T for a
cylindrical one.

\begin{figure}
\begin{center}
\includegraphics*[scale=0.42]{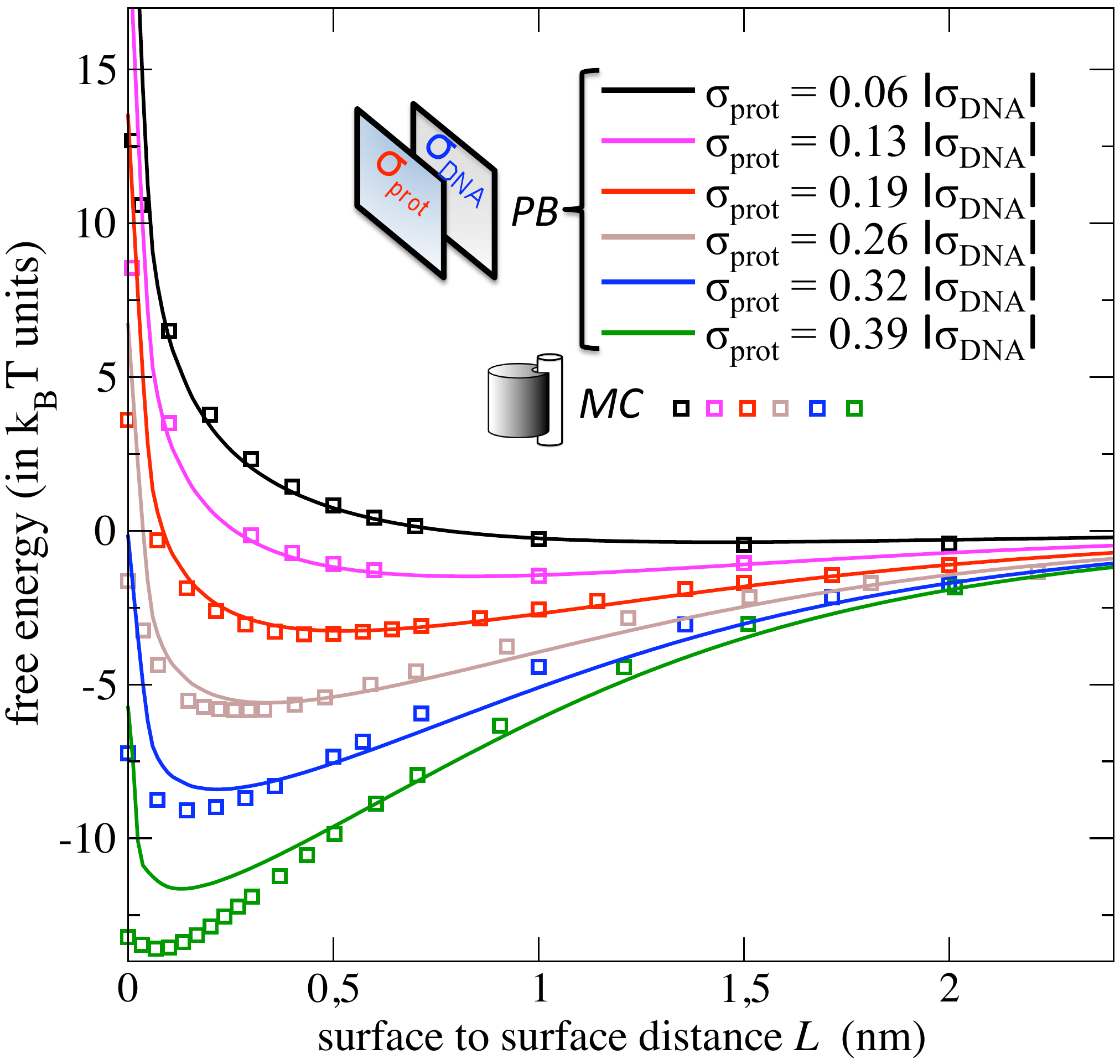}
\caption{{\bf Influence of the protein charge on the interaction.} 
Free energy of the DNA-protein system for a set of protein charge
densities obtained by PB theory (curves) and by MC simulations
(squares). The area of the concave protein surface $S_{\rm int}$ is
$15$~nm$^2$. The charge density is $\sigma_{\rm prot}=Z_{\rm
prot}/S_{\rm int}$, with $Z_{\rm prot}$ the charge of the protein at
the interface. The charge density of DNA is $\sigma_{\rm DNA}=-1.0
\;e$~nm$^{-2}$.  In the MC simulations, the shape model for the DNA-BP
is a cylinder of height $5$~nm, with a concave interface
($C_\perp=-1/R_{\rm DNA}$, $C_\parallel=0$). The protein charges are
distributed on a pattern of 16 sites, $0.1$~nm below the surface of the
cylindrical cavity.
}
\label{fig2}
\end{center}
\end{figure}

Once the role of the protein curvature is established, we perform
simulations of concave DNA-BP models with various charge patterns to
assess the influence of the protein charge on the interaction. When
the pattern changes at constant interface charge density $\sigma_{\rm
prot}$, the free energy exhibits only minor variations (data not shown). 
Conversely, $\sigma_{\rm prot}$ strongly
modulates the free energy profile (Fig.~\ref{fig2}). For an interface
of {\em e.g.} $15$~nm$^2$, if $\sigma_{\rm prot}$ changes from $0.13
\vert{\sigma}_{\rm DNA}\vert$ to $0.39 \vert{\sigma}_{\rm DNA}\vert$,
$F_{\rm min}$ dramatically decreases from $ -2$~k$_{\rm B}$T to $
-14$~k$_{\rm B}$T and $L_{\rm min}$ decreases from $0.75$~nm to
$0.1$~nm.

To provide a rational basis to the simulation results, we carry out
statistical mechanical calculations within the Poisson Boltzmann (PB)
framework. The complementary interacting surfaces of the protein and
the DNA are described by a minimal model: two charged parallel plates
separated by a distance $L$. In agreement with the MC results, this
model predicts a minimum of the free energy, whose depth and position
can be analytically expressed \cite{Ohs75,Pai08}.
Moreover, we introduce corrections to the plate-plate
model to account for the actual curvature of protein and DNA
by rescaling both the interface area $S_{\rm int}$  and the charge density.
More precisely, the PB free energy is
integrated over $S_{\rm int}$ after
projection of each surface element on the plane orthogonal to the $L$
axis \cite{SEI}. If $R$ and $h$ are the radius
and height of the cylindrical interface, the interaction free energy 
is given by
\begin{eqnarray} 
F(L) = 
\int_{-h/2}^{h/2} dx \int_{-R}^{R} dy\; E(L) \sqrt{1-y^2/R^2} = E(L) S_{\rm
int}/{2}
\nonumber \end{eqnarray} 
where $E(L)$ is the interaction free energy by unit area for two parallel
plates and $z$ the distance between two surface elements of the curved
bodies facing each other. The effective charge densities used in the PB calculation
are obtained by fitting 
all the Monte Carlo results simultaneously (${\sigma}_{\rm
DNA}^{\rm eff} \simeq 0.6\; {\sigma}_{\rm DNA}$ and ${\sigma}_{\rm
prot}^{\rm eff} \simeq 1.2\; {\sigma}_{\rm prot}$). 
Despite the nanometer size of the interface, the
Poisson-Boltzmann results remarkably agree with the results of the
Monte Carlo simulations for the concave DNA-BP model
(Fig.~\ref{fig2}).

\begin{figure}[!t]
\begin{center}
\includegraphics*[scale=0.55]{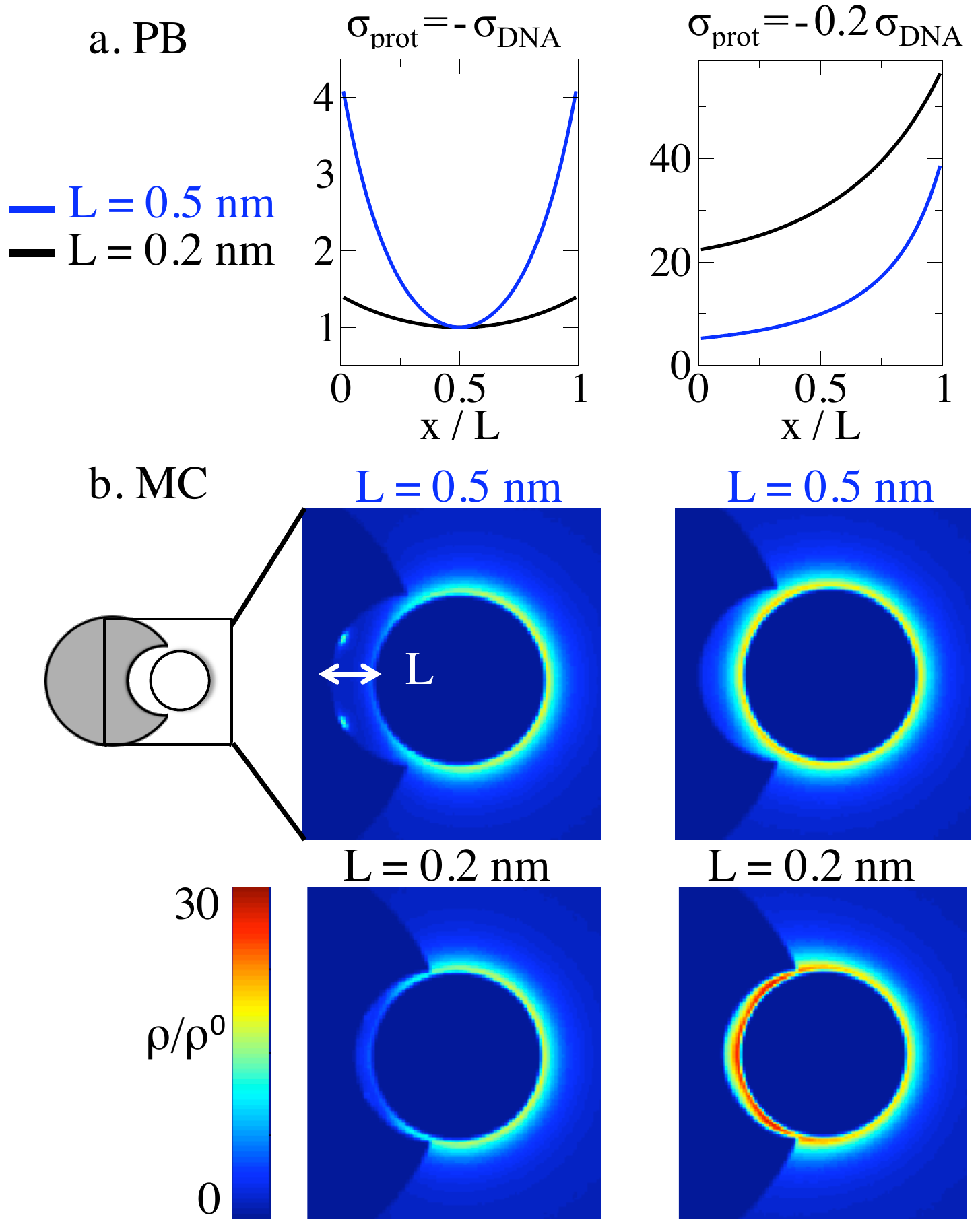}
\caption{{\bf Ionic density fields.} 
The density is obtained by PB theory ({\bf a}), and by MC simulations
({\bf b}) for two protein charge densities and two distances $L$. The
unit is the bulk osmolarity $\rho^0=0.2$~mol.L$^{-1}$. In the PB
treatment, the system is translationally invariant along the
plates. The density along the direction $x$ perpendicular
to the plates is plotted  ($x=0$ on the protein and $x=L$ on DNA). In the
simulations, the DNA-BPs are translationally invariant along the DNA
axis, and the ionic density in the plane perpendicular to the DNA axis is plotted .
}
\label{fig3}
\end{center}
\end{figure}

Furthermore, the PB results shed light on the two physical mechanisms
inducing an attraction and a repulsion between oppositely charged
bodies. The $N_+$ cations and $N_-$ anions between the plates are in
equilibrium with a bulk reservoir ($\mu$VT ensemble). Here, this
equilibrium displays two regimes: a counterion-dominated regime, for
which the number of ions between the plates is dominated by the
counterions neutralizing DNA ($N_+ \gg N_-$), and a salt-dominated regime
($N_+-N_- \ll N_-$). It is established that the salt-dominated regime
is attractive, because the salt release is favorable salt both
entropically (because the volume between the plates decreases) and
electrostatically (because the plates are oppositely charged)
\cite{Ben07}.  As expected, the ionic density decreases as the
charged plates approach each other in the particular case ${\sigma}_{\rm prot} =
\vert{\sigma}_{\rm DNA}\vert$ ({\it i.e.} $N_+ = N_-$) representative
of this regime (Fig.~\ref{fig3}a).  Nevertheless, if ${\sigma}_{\rm
prot}<\vert{\sigma}_{\rm DNA}\vert$, a constant number of neutralizing
counterions remains confined between the plates in order to maintain
electroneutrality.  As $L$ decreases, these cations get more and more
concentrated. Below a given distance, this counterion trapping 
dominates the 
salt release (counterion-dominated regime). As a matter of fact, the ionic density increases as $L$
decreases for ${\sigma}_{\rm prot} = -0.2{\sigma}_{\rm DNA}$,
Fig.~\ref{fig3}a. The resulting enhancement of the osmotic pressure
exceeds the salt-mediated attraction and results in a global
repulsion.

To visualize how this mechanism applies to a more realistic interface,
we compute the ionic density by MC simulations. As shown in
Fig.~\ref{fig3}b, the two regimes are similar to those observed with
the two-plate model. This highlights the significance of
electroneutrality effects for the nanometric interfaces of
biopolymers.  Indeed, since the Debye length $\lambda_D$ ({\em i.e.} the range of
charge inhomogeneities in solution) is of the order of a nanometer,
strong electric fields can appear locally and trap ions in a very
confined space. Moreover, this physical picture explains the influence
of shape complementarity: The interface is then large enough (relative
to $\lambda_D$) and the gap thin enough to trap cations within a small
volume.

To what extent do real DNA-BPs trap ions between their surface and
DNA? To answer this question, 
we perform a statistical analysis of the protein interface charge
densities and complementary surface areas, on a data set of 77
proteins. 
The charge densities of those proteins are not directly available, but
 DNA-BPs are characterized by conserved
propensities of charged residues at the interface region, as defined
in Ref.~\cite{Jon03}. 
For each protein in the data set,
we evaluate the total number of residue $N^{prot}_{tot}$, and the
number of residue $i$ $N^{prot}_i$ for the charged residues ($i$= Arg,
Lys, Asp and Glu). 
Ref.~\cite{Jon03} and Ref.~\cite{Jan99} provides $N^{int}_{tot}$, the
number of residues at the interface.
We estimate the
charge densities of the proteins by approximating the propensity of a residue $i$
by $(N^{int}_i/N^{int}_{tot})/(N^{prot}_i/N^{prot}_{tot})$, and this
leads to the number of residues $i$ at the
interface $N^{int}_i$ and thus the number of charges.
  We take a mean
interface area per residue of $0.70$ ~nm$^2$ \cite{Sta03} to derive the 
mean charge density $\sigma_{\rm prot}$.
 In the case of sequence-specific
DNA-BPs such as transcription factors and restriction enzymes, we
obtain $\sigma_{\rm prot}=(0.17 \pm 0.03) \vert{\sigma}_{\rm
DNA}\vert$. Besides, we notice that the less-specific DNA-BPs
(polymerases, DNA-repair proteins, histones) are more charged
($\sigma_{\rm prot}=(0.27 \pm 0.05)
\vert{\sigma}_{\rm DNA}\vert$).
The area of the fitting interface $S_{\rm prot}=15 \pm 5$ ~nm$^2$ is
similar for all DNA-BPs \cite{Jon99}. According to these structural
features, DNA-BPs should thus be repelled by DNA (cf. Fig.~\ref{fig2}). 
This repulsion obtained with a
coarse-grained model is in agreement with simulations of atomic models
of BamHI \cite{Sun03}, showing a repulsion
when the concave surface of the protein approaches DNA.

To assess whether this repulsion is still significant after addition
of a realistic short-range attraction, we include H-bond interactions
and study the resulting free energy as a function of the protein
position $z$ along the sequence and the distance $L$ between the
surfaces.  We consider a DNA-BP model of charge $\sigma_{\rm
prot}=0.17\vert{\sigma}_{\rm DNA}\vert$ with a fitting shape.
We account for each H-bond
by a Morse potential term $V_M(L) = D [(e^{-\alpha L}-1)^2-1]$ with
$D=0.5$~k$_{\rm B}$T \cite{Tar02} and $\alpha=20$ nm$^{-1}$ \cite{Che04}.  
Crystal structures of protein-DNA
complexes provide a value of the number of H-bonds $n_{\rm spec}$ at
the specific site ($30$ H-bonds for
 $S_{\rm int} = 20$
nm$^2$ \cite{Jan99}). We assume that the number $n$ of
H-bonds that the protein can make on non-specific DNA follows a Gaussian
distribution of average $\langle n \rangle=n_{\rm spec} /3$, and standard
deviation $\sigma_n=\sqrt{n_{\rm spec}}$.
% memo:
% Jones : 1.5 Hbonds/100A2 in average, with an average interaction surface of about 2000 A2,
% this leading to (1.5/100)*3000 = 30 bonds. 
% nb: 15 Hbonds et 15 hydrophobic in Kalodimos and 18 Hbonds for ecoRV
The value of  $\langle n \rangle$ is low
because the number of H-bonds dramatically
decreases for non specific sequences, even for sequences with a high degree of
homology to the target one, as observed in the crystal structure of non cognate BamHI
complex in Ref.~\cite{Via00}. 

\begin{figure}[!t]
\begin{center}
\includegraphics*[width=0.5\textwidth]{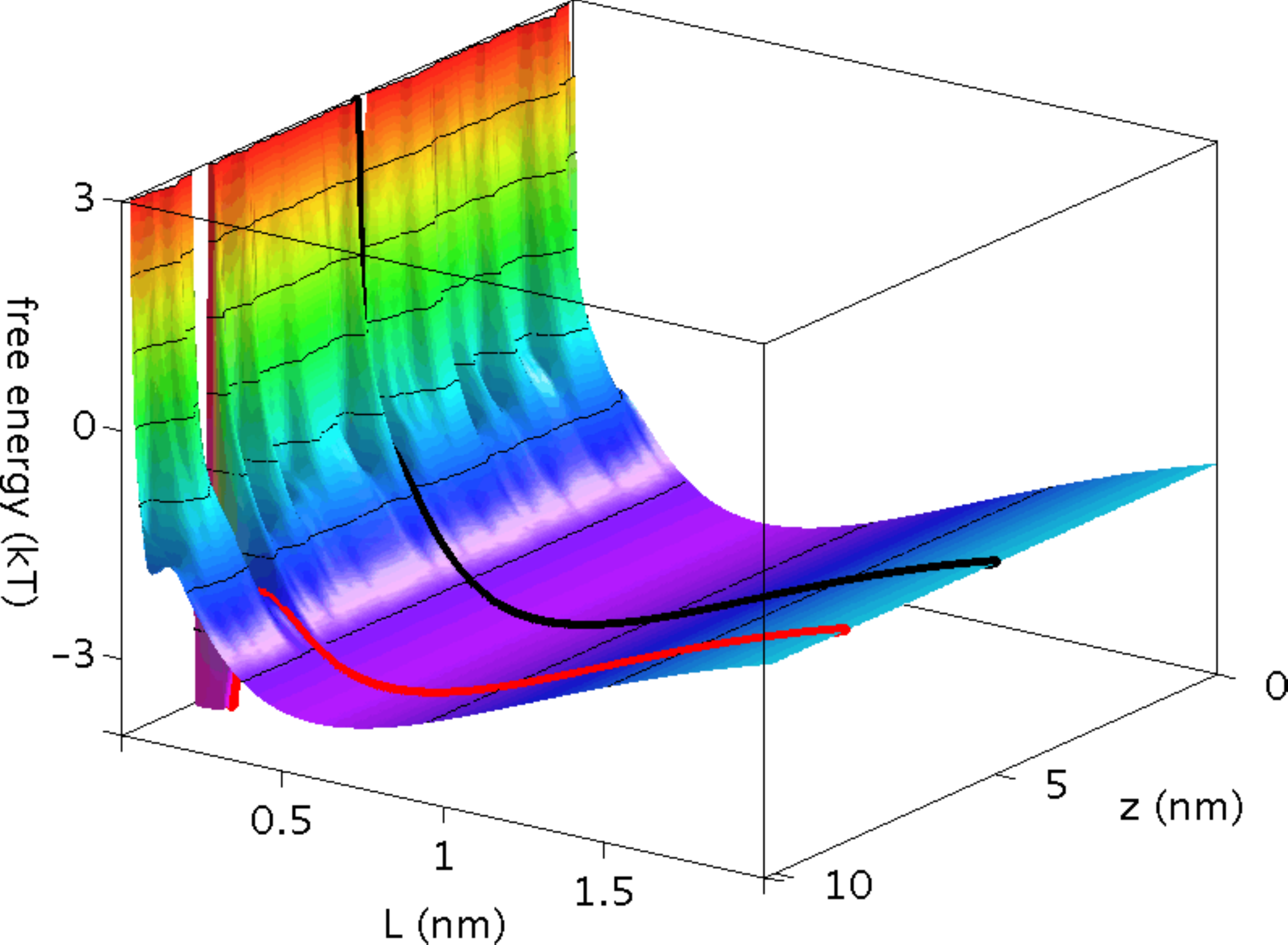} \\
\includegraphics*[width=0.44\textwidth]{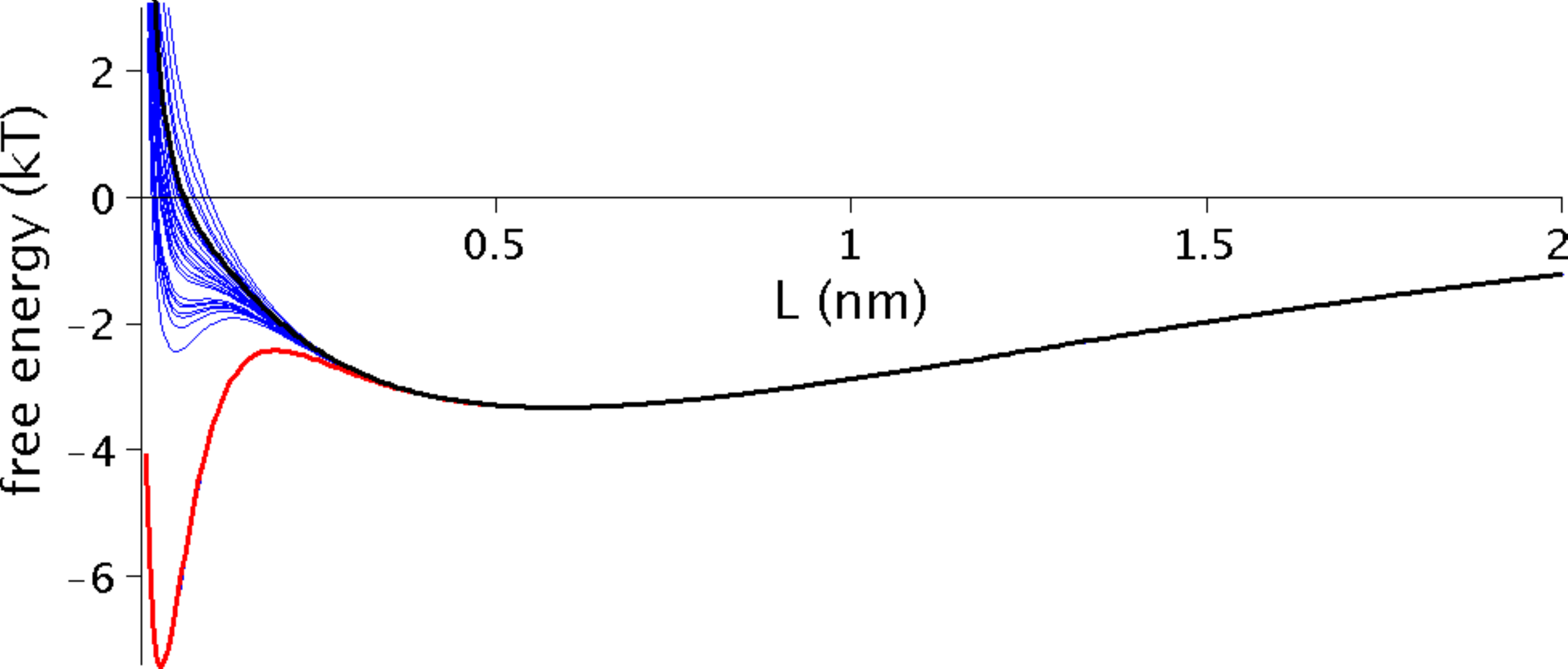}
\caption{{\bf Free energy landscape.}
The free energy is computed along a 30 bp DNA sequence, as a function
of $L$ and of the protein coordinates along DNA ($z$), for
$\sigma_{\rm prot}=0.17\vert{\sigma}_{\rm DNA}\vert$.  The gap between
level lines is $k_BT$.  For more clarity, the additional lower graph
displays the free energy as a function of $L$ for each $z$ value. In
both graphs, the black curve corresponds to a randomly chosen
non-specific coordinate, while the red curve corresponds to the
specific-site. }
\label{fig4}
\end{center}
\end{figure}

The resulting free energy landscape is shown in
Fig.~\ref{fig4}. Remarkably, the osmotic repulsion between
sequence-specific DNA-BPs and DNA dominates along non-specific
sequences.  The equilibrium gap distance of nearly
$0.5$~nm is in agreement with the distance observed in the complexes
of EcoRV ($0.51$~nm \cite{Jon99}) with non-specific sequences.
Interestingly, along the equilibrium valley, the
roughness of the sequence-dependent part of the potential is screened
out: The protein can therefore easily slide along DNA. At the target
site, the large H-bond interaction significantly reduces the barrier, and the
protein can approach the DNA.

%Our results unravel a subtle balance between long-range electrostatic
%attraction, short-range osmotic repulsion and short-range attraction.
%This effect is sensitive to the shape and charge of DNA-BPs, and
%should have thus contributed to the structural evolution of those
%proteins.  It also provides a switching mechanism  between a
%searching mode on non-specific sequences and a recognition mode at the
%sequence. The implications of such a behavior on the protein 1D
%diffusion along DNA recently observed both {\em in vitro} and {\em in
%vivo} \cite{Gow05,Elf07,Deb08,revue2008} could be relevant and
%will be the goal of future investigations.
Our results unravel a subtle balance between long-range electrostatic
attraction, short-range osmotic repulsion and short-range attraction.
This effect is sensitive to the shape and charge of DNA-BPs, and
should have thus contributed to the structural evolution of those
proteins.  From a dynamical perspective, our model provides new bases
to conciliate the dual requirement of high protein mobility and high
sequence sensitivity \cite{Klafter, Metzler,BeniCondmatt}. Indeed, the latter is
usually assumed to slow down the protein diffusion
\cite{Slu04,Bar04a}. According to our results, the DNA-BP freely
diffuses along non-specific DNA, confined in an electrostatic free
energy valley.  The free energy barrier, which keeps the protein at a
distance from DNA, is also a signature of the sequence: Transverse
thermal fluctuations enable the protein to cross the barrier only at
the specific site or at highly homologous sequences.  This recognition
mechanism is efficient because it does not require the protein to
probe the molecular details of non-specific DNA sequences.  The
implications of such a behavior on the protein 1D diffusion along DNA
recently observed both {\em in vitro} and {\em in vivo}
\cite{Gow05,Elf07,Deb08,revue2008} will be the goal of future
investigations.

\end{document}